\documentclass[reprint, amsmath, amssymb, aps, superscriptaddress, prx]{revtex4-2}

\usepackage{bm,color,amsmath,txfonts}

\usepackage{graphicx}
\usepackage{subfigure}
\usepackage{verbatim}
\usepackage{dcolumn}
\usepackage{bm}
\usepackage{epsf}
\usepackage{xcolor}
\usepackage{hyperref}
\usepackage{hhline}
\usepackage{float}
\usepackage{enumerate}
\usepackage{bbm}
\usepackage{lipsum}
\usepackage{natbib}
\usepackage{soul}
\definecolor{mygreen}{rgb}{0.01, 0.31, 0.59}
\definecolor{myblue}{rgb}{0.01, 0.31, 0.59}
\definecolor{myred}{rgb}{0.63, 0.12, 0.12}
\hypersetup{
	colorlinks=true,   
	linkcolor=blue,  
	citecolor=blue, 
	urlcolor =blue    
}

\hypersetup{
	colorlinks=true,   
	linkcolor=blue,  
	citecolor=blue, 
	urlcolor =blue    
}

\renewcommand{\thefigure}{\textbf{\arabic{figure}}}
\usepackage{ulem}

\begin{document}
	
\title{Polaromechanics: cavity-magnon polaritons strongly coupled to phonons}

\author{Rui-Chang Shen}
\affiliation{Zhejiang Key Laboratory of Micro-Nano Quantum Chips and Quantum Control, School of Physics, and State Key Laboratory for Extreme Photonics and Instrumentation, Zhejiang University, Hangzhou 310027, China}
\author{Jie Li}\email{Corresponding author: jieli007@zju.edu.cn}
\affiliation{Zhejiang Key Laboratory of Micro-Nano Quantum Chips and Quantum Control, School of Physics, and State Key Laboratory for Extreme Photonics and Instrumentation, Zhejiang University, Hangzhou 310027, China}
\author{Yi-Ming Sun}
\affiliation{Zhejiang Key Laboratory of Micro-Nano Quantum Chips and Quantum Control, School of Physics, and State Key Laboratory for Extreme Photonics and Instrumentation, Zhejiang University, Hangzhou 310027, China}
\author{Wei-Jiang Wu}
\affiliation{Zhejiang Key Laboratory of Micro-Nano Quantum Chips and Quantum Control, School of Physics, and State Key Laboratory for Extreme Photonics and Instrumentation, Zhejiang University, Hangzhou 310027, China}
\author{Xuan Zuo}
\affiliation{Zhejiang Key Laboratory of Micro-Nano Quantum Chips and Quantum Control, School of Physics, and State Key Laboratory for Extreme Photonics and Instrumentation, Zhejiang University, Hangzhou 310027, China}
\author{Yi-Pu Wang}
\affiliation{Zhejiang Key Laboratory of Micro-Nano Quantum Chips and Quantum Control, School of Physics, and State Key Laboratory for Extreme Photonics and Instrumentation, Zhejiang University, Hangzhou 310027, China}
\author{Shi-Yao Zhu}
\affiliation{Zhejiang Key Laboratory of Micro-Nano Quantum Chips and Quantum Control, School of Physics, and State Key Laboratory for Extreme Photonics and Instrumentation, Zhejiang University, Hangzhou 310027, China}
\affiliation{College of Optical Science and Engineering, Zhejiang University, Hangzhou 310027, China}
\affiliation{Hefei National Laboratory, Hefei 230088, China}
\author{J. Q. You}\email{Corresponding author: jqyou@zju.edu.cn}
\affiliation{Zhejiang Key Laboratory of Micro-Nano Quantum Chips and Quantum Control, School of Physics, and State Key Laboratory for Extreme Photonics and Instrumentation, Zhejiang University, Hangzhou 310027, China}
\affiliation{College of Optical Science and Engineering, Zhejiang University, Hangzhou 310027, China}

\maketitle

{\bf \noindent Building hybrid quantum systems is a crucial step for realizing multifunctional quantum technologies, quantum information processing, and hybrid quantum networks. A functional hybrid quantum system requires strong coupling among its components. However, couplings between distinct physical systems are typically very weak. Experimental realization of strong coupling in a hybrid system remains a long-standing challenge, especially when it has multiple components and the components are of different nature.  Here we demonstrate the realization of triple strong coupling in a novel {\it polaromechanical} hybrid system, where polaritons, formed by strongly coupled ferromagnetic magnons and microwave photons, are further strongly coupled to phonons. The corresponding polaromechanical normal-mode splitting is observed. A high polaromechanical cooperativity of $9.4\times10^3$ is achieved by significantly reducing the polariton decay rate via exploiting coherent perfect absorption. The quantum cooperativity much greater than unity is achievable if placing the system at cryogenic temperatures, which would enable various quantum applications. Our results pave the way towards coherent quantum control of photons, magnons and phonons, and are a crucial step for building functional hybrid quantum systems based on magnons. }

\vspace{0.3cm}
\noindent Quantum information processing, quantum technologies, and the building of hybrid quantum networks require hybrid quantum systems (HQSs), which combine different physical systems with their individual strengths and complementary functionalities, to store, process, and transmit quantum information~\cite{You13,Rabl15,Clerk20,Smith16}. A prerequisite for realizing a functional HQS is the ability to exchange quantum states between its components with high fidelity, which requires strong coupling, 
where reversible energy exchange between interacting systems is faster than their energy dissipation to the environment.  To date, strong coupling has been realized in a variety of quantum systems, such as atoms and optical photons~\cite{Kimble1992}, superconducting qubits and microwave photons~\cite{Schoelkopf2004}, excitons and microcavity photons~\cite{Weisbuch1992}, among others.  In terms of mechanical systems, promising components for HQSs given their ability to couple with various quantum systems and high quality factors, strong coupling has been achieved with optical or microwave photons~\cite{Simon2009,Teufei2011,Kippenberg2012}, a superconducting qubit~\cite{Cleland2010,Chu17,Cleland2018,Chu18,Painter20}, a quantum dot~\cite{Clerk2010}, atomic spins~\cite{Treutlein2020}, etc. This enables not only the function of a mechanical ``quantum bus" for interfacing and communicating between different quantum systems, but also the preparation and control of quantum states of mechanical oscillators, which has been a subject of long-standing interest~\cite{Schwab}.

Of particular interest is to build HQSs composed of more than two components, ranging from photons, atoms, and spins to superconducting and nanomechanical structures, especially when they are of different nature, where hybridization is maximized to integrate complementary functionalities. Over the past few years, collective spin excitations (magnons) in magnetic materials, e.g. yttrium iron garnet (YIG), have been demonstrated as possible building blocks for novel quantum technologies~\cite{Naka19,Awschalom21}, quantum information processing~\cite{Liy20}, quantum networks~\cite{Li21}, and quantum computing~\cite{Chumak22}. Their unique feature is that they can coherently interact with a variety of quantum systems including microwave or optical photons, superconducting qubits, {acoustical} phonons, etc~\cite{Yuan,Bauer22,Pirro}.   However, couplings between excitations of distinct physical systems are typically very weak, and this is particularly the case when the excitation frequencies differ greatly~\cite{Rabl15}. {Impressively, the new quasiparticles termed phonoritons -- arising from the strong coupling between photons, phonons, and excitons, have been predicted~\cite{Latini2021} and recently observed~\cite{Kuznetsov2023} in a microcavity.  However, up to now, the realization of such a triple strong coupling in hybrid magnonic systems} remains challenging and an outstanding goal.


Here, {we report an experimental realization of strong coupling in a novel  polaromechanical system~\cite{Santos}, which involves three different quanta, namely, ferromagnetic magnons, microwave photons, and long-lived phonons.  In the experiment, magnon polaritons (MPs)~\cite{Basov}, formed by strong interaction between magnons and microwave cavity photons, are further strongly coupled with vibrational phonons that are induced by the magnetostriction of a YIG sphere.}   The decay rate of the MP can be significantly reduced by operating the system at the conditions of coherent perfect absorption (CPA)~\cite{Chong2010,Cao2011,Cao2012,Yang2021}, under which the microwave cavity mode achieves an effective gain~\cite{Zhang17}. By further operating the system at the (cavity) gain-(magnon) loss balance, the decay rate of the MP is reduced to nearly zero. This, together with the small mechanical damping rate, leads to the strong coupling between the mechanical mode and the driven MP, as witnessed by an evident normal-mode splitting (NMS) at the mechanical sideband.  The normal modes of the system are then the hybridization of photons, magnons and phonons, {which explicitly indicates that the system is in the triple strong-coupling regime.} A remarkably high polaromechanical cooperativity $\sim 9.4\times10^3$ is accordingly achieved.

\begin{figure}[t]
\includegraphics[width=\linewidth]{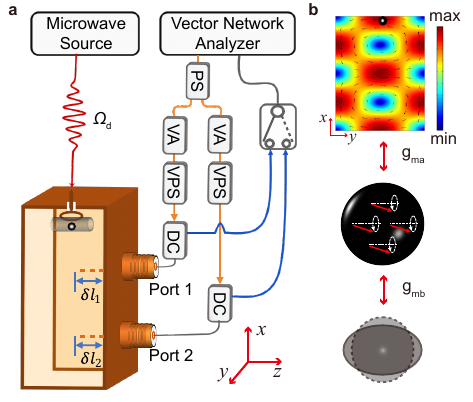}
\caption{\textbf{Schematic diagram of the experimental setup.
a} The microwave signal generated by a vector network analyzer (VNA) is equally divided into two beams through a power splitter (PS). The amplitude and phase of each beam can be independently tuned by a variable attenuator (VA) and variable phase shifter (VPS). The two microwave fields are injected into the cavity after through two directional couplers (DCs) connected to the two ports, and the output fields are measured by the VNA after the DCs. The microwave cavity is a 3D oxygen-free copper cavity with dimensions of $60\times25\times6~{\rm{mm}^{3}}$, and a 0.25-mm-diameter YIG sphere is placed at the antinode of the magnetic field of the cavity $\rm{TE}_{102}$ mode. The YIG sphere is supported by a horizontal glass capillary and can move freely to reduce the mechanical damping. At the top of the cavity, a coil connected with a microwave source is used to directly drive the YIG sphere.  Each cavity port is fixed with a SMA adaptor, and the length of the SMA adaptor's pin entering the cavity $\delta l_{1(2)}$ can be tuned. The external magnetic field $B_0$ (in the $z$ direction) is applied to set the frequency of the magnon mode. The experiment is performed at room temperature. \textbf{b} Three interacting modes of the system, including a cavity mode, a magnon mode, and a vibration mode. At the top is the magnetic field distribution of the cavity $\rm{TE}_{102}$ mode.}
\label{fig1}
\end{figure}

\vspace{0.6cm}
\noindent\textbf{Results}\\
\noindent\textbf{The polaromechanical system}\\
\noindent
The polaromechanical system is schematically shown in Fig.~\ref{fig1}a, which consists of a 3D microwave cavity and a YIG sphere supporting both a magnon mode and a deformation vibration mode~\cite{Zuo24}. The magnon mode couples to the microwave cavity mode via the magnetic dipole interaction~\cite{Huebl2013,Nakamura2014,Tang2014}, and to the mechanical vibration mode via the magnetostrictive interaction~\cite{Tang,Jie18,Davis,Li2022} (Fig.~\ref{fig1}b).  The cavity $\rm{TE_{102}}$ mode has a resonance frequency $\omega_{\rm{a}}/2\pi=7.213~\rm{GHz}$ and a total decay rate $\kappa_{\rm{a}}$, which is contributed by three dissipation channels, i.e., $\kappa_{\rm{a}}=\kappa_{\rm{int}}+\kappa_{\rm{1}}+\kappa_{\rm{2}}$, with the intrinsic decay rate $\kappa_{\rm{int}}/2\pi=1.98~\rm{MHz}$, and the external decay rate $\kappa_{\rm{1(2)}}$ due to the connection to the cavity port 1(2). To implement the CPA, $\kappa_{\rm{1(2)}}$ is set to be tunable by adjusting the length $\delta l_{1(2)}$ of the SMA adaptor's pin into the cavity.
The frequency of the magnon mode can be continuously adjusted by varying the external bias magnetic field $B_{0}$ via $\omega_{\rm m}=\gamma B_{0}$, with the gyromagnetic ratio $\gamma/2\pi=28~{\rm GHz/T}$. The magnon decay rate $\kappa_{\rm{m}}/2\pi=0.49~\rm{MHz}$, and the magnon-cavity coupling strength $g_{\rm{ma}}/2\pi=6.63~\rm{MHz}$.  The magnon mode is driven by a microwave field loaded via a loop antenna, which further drives the mechanical mode through the magnomechanical coupling.   In the experiment, we observe three mechanical modes with close resonance frequencies~(see the Supplementary Information (SI) for details), and we focus on the mode of frequency $\omega_{\rm{b}}/2\pi=10.9565~\rm{MHz}$, which exhibits the strongest bare magnomechanical coupling rate $g_{\rm{mb}}/2\pi=1.40~\rm{mHz}$ and the lowest damping rate $\kappa_{\rm{b}}/2\pi=155~\rm{Hz}$.

Thanks to the high spin density of the YIG, the magnon mode and the cavity are strongly coupled, $g_{\rm{ma}} > \kappa_{\rm{a}}, \kappa_{\rm{m}}$, leading to two hybridized MPs~\cite{Huebl2013,Nakamura2014,Tang2014}.  The corresponding frequencies $\omega_{\pm}$ and decay rates $\kappa_{\pm}$ of the two polariton modes are given by $\omega_{\pm}-i\kappa_{\pm}= \frac{1}{2}\left[ \left(\omega_{\rm{a}}+\omega_{\rm{m}}\right) -i\left(\kappa_{\rm{a}}+\kappa_{\rm{m}}\right) \pm \xi^{\frac{1}{2}} \right]$, where $\xi= 4g_{\rm{ma}}^2 +\left[ \left(\omega_{\rm{a}}-\omega_{\rm{m}}\right) -i\left(\kappa_{\rm{a}}-\kappa_{\rm{m}}\right)\right]^2$.  The linearized Hamiltonian of the system in terms of two polariton modes reads~(see SI)
\begin{equation}\label{f-1}
\begin{aligned}
H/\hbar =& \omega_{\rm{+}} p^{\dagger}_{+} p_{+}+\omega_{\rm{-}} p^{\dagger}_{-} p_{-}+\omega_{\rm{b}} b^{\dagger} b\\
		&+\left(G_{+} p_{+}^\dagger+ G_{+}^{*} p_{+} \right)x +\left(G_{-} p_{-}^\dagger+ G_{-}^{*} p_{-} \right) x ,
\end{aligned}
\end{equation}
where $p_{+}=a\cos\theta + m\sin \theta$ and $p_{-}=-a\sin \theta + m\cos \theta$ denote the annihilation operators of the upper- and lower-branch polaritons, respectively, satisfying the bosonic commutation relation $[p_{\pm}, p_{\pm}^\dag ]=1$, and $\theta= \frac{1}{2} \arctan \left(\frac{2g_{\rm {ma}}}{\omega_{\rm{a}}-\omega_{\rm{m}}} \right)$ determines the proportion of the magnon (cavity) mode in the two polaritons; 
$a$, $m$, and $b$ ($a^{\dag}$, $m^{\dag}$, and $b^{\dag}$) are the annihilation (creation) operators of the cavity, magnon and mechanical modes, respectively, and $x= \left(b+ b^\dagger\right)/\sqrt{2}$ denotes the mechanical displacement by magnetostriction; $G_{+(-)}$ is the effective coupling strength between the upper (lower)-branch polariton and the mechanical mode, and their expressions are
\begin{equation}\label{f-2}
\begin{aligned}
G_{+} =\sqrt{2}g_{\rm{mb}} M\sin \theta,~\,\,\,\,
G_{-} =\sqrt{2}g_{\rm{mb}} M\cos \theta ,
\end{aligned}
\end{equation}
where $M \equiv  \langle m\rangle$ is the steady-state average of the magnon mode and $n_{\rm mag}=|M|^2$ is the number of magnon excitations. {A similar polaromechanical coupling exists in an exciton-photon-phonon system~\cite{Sesin23}.}
Equation \eqref{f-2} indicates that the polaromechanical  coupling $G_{+(-)}$ can be significantly enhanced by strongly driving the magnon mode. When the coupling rate exceeds the dissipation rates of the polariton and mechanical modes, i.e., $G_{+(-)} > \kappa_{+(-)}, \kappa_{\rm b}$, the system enters the {polaromechanical} strong-coupling regime, in which the normal modes are the hybridization of photons, magnons and phonons.  Previous experiments have demonstrated relatively weak couplings, i.e., $G_{\pm}$ up to tens of kHz~\cite{Tang,Davis,Li2022}, due to the intrinsically small bare coupling $g_{\rm{mb}}\,{<}\,10$ mHz, which is much smaller than the polariton decay rates $\kappa_{\pm} \sim 1$~MHz (limited by the magnon intrinsic loss). The enhancement of the coupling strength $G_{\pm}$ by further increasing the drive power is restricted by strong nonlinear effects of the YIG sphere \cite{Li2022,Suhl1955}. Therefore, the {polaromechanical strong coupling} of the system is yet to be achieved.

\begin{figure*}[t]
\includegraphics[width=\linewidth]{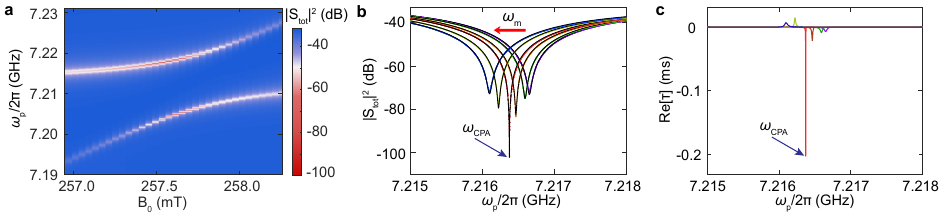}
\caption{\textbf{Reduction of the polariton decay rate and characterization of the CPA.
a} Output spectrum of the CPA with $\kappa_{\rm{1}}/2\pi=1.22~\rm{MHz}$, $\kappa_{\rm{2}}/2\pi=1.24~\rm{MHz}$, $\Delta\phi=0$,  and $q\approx 1$. {The output spectrum $|S_{\rm tot}|^2=|r_i+t_j|^2$ ($i\neq j=1,2$) contains the reflection (transmission) field of the input field loaded at port $i$ ($j$).} $\omega_{\rm p}$ is the frequency of the input field. The frequency of the magnon mode is tuned by varying the bias magnetic field {$B_0$}. 
\textbf{b} Output spectra of the upper-branch polariton with different magnon frequencies. {As the frequency of the polariton approaches the CPA frequency $\omega_{\rm CPA}$, via tuning the magnon frequency ({which is tuned from 7.2116 to 7.2104 GHz by reducing $B_0$ from 257.55 to 257.51 mT, as indicated by the red arrow}), the polariton decay rate decreases manifesting as a rapid reduction of the output field. The smallest decay rate of the polariton is achieved at the CPA frequency.}  Black curves are fitting results. \textbf{c} The Wigner time delay ${\rm Re}[\tau(\omega)]$ corresponding to the output spectra in \textbf{b}. The long time delay $\sim0.2$~ms signifies the significantly reduced decay rate of the polariton at the CPA frequency. }
\label{fig2}
\end{figure*}

\vspace{0.3cm}
\noindent\textbf{Polariton decay rate reduction via CPA}\\
\noindent
A natural idea to achieve the polaromechanical  strong coupling is to reduce the dissipation rate of the polariton mode. To this end, we exploit the CPA to achieve an effective gain of the cavity mode. {The CPA refers to the fact that, due to the destructive interference between the input and output fields at the cavity port, the amplitude of the cavity output field turns out to be zero~\cite{Chong2010,Cao2011,Cao2012,Yang2021}.} This approach has been adopted to realize the exceptional point associated with the parity-time symmetry in cavity magnonics~\cite{Zhang17} and optical microcavities~\cite{Yang2021}.   Under the CPA conditions, the output field of the cavity $A^{\rm{out}}_{1(2)}=0$. The input-output relation $A^{\rm{out}}_{1(2)}=\sqrt{2\kappa_{\rm{1(2)}}}A-a^{\rm{in}}_{1(2)}$~\cite{Collett1985} thus leads to the input field $a^{\rm{in}}_{1(2)}= \sqrt{2\kappa_{\rm{1(2)}}}A$, which provides a gain for the cavity mode (see SI). The cavity mode is then compensated to be a gain mode with an effective gain rate~\cite{Zhang17}: $\kappa_{\rm a}' \equiv \kappa_{\rm 1} +\kappa_{\rm 2} - \kappa_{\rm int}$. The CPA requires that the cavity gain and the magnon loss are  balanced, i.e., $\kappa_{\rm a}' = \kappa_{\rm m}$, for the resonant case of $\omega_{\rm m}=\omega_{\rm a}$, which leads to the vanishing of the polariton decay rates, i.e., $\kappa_{\pm}=\frac{-\kappa_{\rm a}' +\kappa_{\rm m}}{2}=0$~(see SI for derivation).  The significantly reduced polariton decay rates, together with the intrinsically small mechanical damping rate, offers the possibility to observe the polaromechanical strong coupling.

In the experiment, accurate realization of the gain-loss balance is vital to obtain vanishing polariton decay rates. To this end, we alter the length of the pins $\delta l_{1(2)}$ to vary $\kappa_{\rm 1(2)}$ in order to meet $\kappa_{\rm a}' = \kappa_{\rm m}$. Figure~\ref{fig2}a shows the output spectrum with $\kappa_{\rm{1}}/2\pi=1.22~\rm{MHz}$ and $\kappa_{\rm{2}}/2\pi=1.24~\rm{MHz}$. The two input fields are adjusted to have a zero phase difference $\Delta\phi=0$ and a power ratio $q=\kappa_{2}/\kappa_{1}\approx 1$~(see SI).  The phase difference and power ratio are tuned via the VPS and VA connected to each port (Fig.~\ref{fig1}a). As shown in Fig.~\ref{fig2}a, when the frequency of the magnon mode approaches the cavity resonance, the output spectrum exhibits remarkable dips. Due to the monochromaticity of the CPA~\cite{Chong2010,Cao2012,Davy2021}, the output spectrum is nearly zero at the CPA frequency $\omega_{\rm CPA}$. Consequently, we obtain the smallest symmetric decay rates of the two MPs $\kappa_{\pm}/2\pi=7.75~\rm{kHz}$, limited by the regulation precision.

To further reduce the decay rate of the polariton, we tune the magnon frequency by varying the bias magnetic field while keeping the cavity frequency fixed. This further reduces the decay rate of one polariton at the price of increasing that of the other~(see SI for details)~\cite{Sermage1996}.  In this way, by red-shifting the magnon frequency we achieve the decay rate of the upper-branch polariton $\kappa_{+}/2\pi=0.78$ kHz, corresponding to an output of $-100.4$~dB (Fig.~\ref{fig2}b). 

The significantly reduced polariton decay rate can be substantiated by calculating the Wigner time delay (WTD)~\cite{Davy2021,Chen2021}, which characterizes the trapped time of the input field in the resonator (i.e., the MP).   The CPA corresponds to the input field being infinitely trapped in the resonator, which can be translated into a diverging WTD. The exact CPA is unmeasurable in the experiment: one can only achieve nearly perfect absorption by improving the control accuracy. This corresponds to a finite WTD, which can be used to estimate the decay rate of the polariton.  The WTD is defined as the real part of the complex-valued $\tau(\omega)$ \cite{Davy2021,Chen2021}, i.e.,
\begin{equation}\label{TD}
\tau(\omega)= \frac{-i\psi_{\rm in}^{*}S^{*}(\omega)\partial_{\omega}S(\omega)\psi_{\rm in}}{\psi_{\rm in}^{*}S^{*}(\omega)S(\omega)\psi_{\rm in}},
\end{equation}
where $S(\omega)$ is the scattering matrix~(see SI for details) and $\psi_{\rm in}=\left(a^{\rm{in}}_{1}, a^{\rm{in}}_{2}\right)^{\rm T}$ is the vector of input fields. In Fig.~\ref{fig2}c, we calculate the WTD associated with the output spectra in Fig.~\ref{fig2}b. Note that there is a one-to-one correspondence between the spectra in Fig.~\ref{fig2}b and the WTD in Fig.~\ref{fig2}c {\it at each frequency}, and a phase transition of ${\rm Re}[\tau]$ occurs at the singularity point, i.e., the exact CPA~\cite{Davy2021}. 
At the CPA frequency, we obtain a maximum WTD $ |{\rm Re} \left[ \tau(\omega_{\rm CPA})\right]| =0.2~{\rm ms}$, corresponding to the decay rate of the polariton $\kappa_{+}=1/{\rm Re}[\tau(\omega_{\rm CPA})]=2\pi\times 0.79~{\rm kHz}$. This agrees well with the value $\kappa_{+}/2\pi=0.78$ kHz evaluated using the theory provided in the SI.

\begin{figure*}[t]
	\hskip-0.19cm\includegraphics[width=0.68\linewidth]{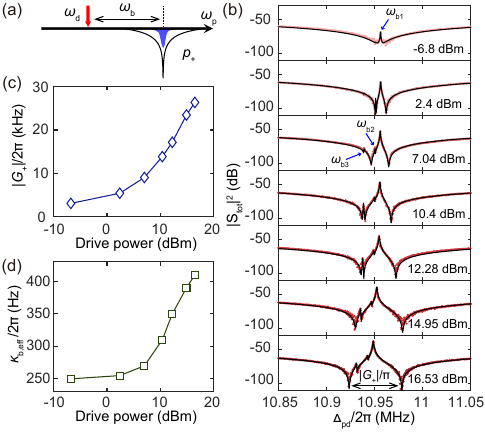}
	\caption{\textbf{Polaromechanical normal-mode splitting. a} Frequency schematic diagram associated with the measurement. {A red-detuned microwave field (red arrow) is used to drive the upper-branch polariton. The detuning is equal to the mechanical frequency, such that the scattered anti-Stokes mechanical sideband is in resonance with the polariton.} \textbf{b} Output spectra of the polaromechanical  normal modes for different drive powers, with $\Delta_{\rm{pd}}=\omega_{\rm p}-\omega_{\rm{d}}$. The corresponding decay rates of the upper-branch polariton $\kappa_{+}/2\pi$ are 6.99, 0.25, 1.26, 1.45, 1.42, 1.50 and 0.45 kHz, respectively. We observe three adjacent mechanical modes and focus on the normal-mode splitting associated with the mode of frequency $\omega_{\rm{b1}}/2\pi=10.9565~\rm{MHz}$~(see SI for more details). {The other two modes of frequencies $\omega_{\rm{b2}}$ and $\omega_{\rm{b3}}$ are identified in the figure}. Black solid curves are fitting results.  \textbf{c} The pump-enhanced effective polaromechanical coupling strength $|G_{+}|$ versus drive power.  \textbf{d} Effective mechanical damping rate $\kappa_{\rm b,eff} $ versus drive power.}
	\label{fig3}
\end{figure*}

\vspace{0.3cm}
\noindent\textbf{Polaromechanical normal-mode splitting}\\
\noindent
The coupling between the MPs and the mechanical mode essentially originates from the dispersive magnetostrictive (magnon-phonon) interaction, which leads to the linearized Hamiltonian in equation~\eqref{f-1}. It reveals that the interaction between each polariton and the mechanical mode is analogous to the linearized opto-~\cite{MAreview} and magnomechanical~\cite{Jie18} Hamiltonian.  This promises similar polaromechanical backaction as the opto- and magnomechanical ones, such as mechanical cooling and amplification~\cite{OMrmp,Zuo24}. Equation \eqref{f-2} indicates that the effective polaromechanical coupling $G_{\pm}$ can be enhanced by increasing the magnon excitation number {$n_{\rm mag}$}. In the experiment, we adopt a red-detuned microwave field to drive the upper-branch polariton and to keep the scattered mechanical sideband in resonance with the polariton (Fig.~\ref{fig3}a).  This enhances the magnomechanical anti-Stokes scattering, which cools the mechanical motion, yielding an increased mechanical damping rate~\cite{Davis}. 

In the first graph of Fig.~\ref{fig3}b, a low drive power of $-6.8$~dBm leads to a weak coupling $|G_{+}|/2\pi = 3.06$ kHz, and the system is in the weak-coupling regime because $|G_{+}| <\kappa_{+} = 2\pi \times 6.99$~kHz. The anti-Stokes sideband is manifested as the magnomechanically induced transparency~\cite{Tang}, and the mechanical damping rate is increased from $\kappa_{\rm b}/2\pi=155$ Hz to $\kappa_{\rm b,eff}/2\pi = 250$ Hz.
By further increasing the drive power, $G_+$ becomes greater than both $\kappa_{+}$ and $ \kappa_{\rm b,eff}$. For example in the second graph, $|G_{+}|/2\pi=5.40~\rm{kHz}$ at the power of $2.4$ dBm, which is greater than $\kappa_{+}/2\pi \,{=}\, 0.25~\rm{kHz}$ and $\kappa_{\rm b,eff}/2\pi \,{=}\, 255~\rm{Hz}$. The system then enters the {polaromechanical strong-coupling} regime.  
From the top to the bottom graph of Fig.~\ref{fig3}b, as the power increases from $-6.8$~dBm to $16.53$~dBm, $|G_+|$ gradually increases to $25.71~\rm{kHz}$ (corresponding to a maximum polaromechanical cooperativity $C_{\rm +,b} \equiv |G_+|^2/(\kappa_+ \kappa_{\rm b}) \approx 9.40\times 10^3$), which exhibits a more evident feature of the NMS with the splitting of $2|G_+|$ in the spectra.   Similar strong coupling induced NMS has been observed in cavity optomechanics~\cite{Simon2009,Teufei2011,Kippenberg2012}. Note that in Fig~\ref{fig3}b, the frequency of the mechanical mode slightly decreases when increasing the power, which is due to the magnon-phonon cross-Kerr effect~\cite{Li2022} (see SI for more details). {In addition, in the graphs under the drive power of 10.4 and 12.28 dBm, we observe the emergence of another weaker, yet distinct normal-mode splitting, which occurs when the polaromechanical strong coupling (associated with the mechanical mode of frequency $\omega_{\rm b1}$) induced normal mode resonates with the mechanical sideband associated with the mechanical mode of frequency $\omega_{\rm b3}$.}  Figures~\ref{fig3}c and \ref{fig3}d show the corresponding coupling strength $|G_+|$ and the effective mechanical damping rate $\kappa_{\rm b,eff}$ for the drive powers used in Fig.~\ref{fig3}b.

\begin{figure}[t]
\hskip-0.19cm\includegraphics[width=\linewidth]{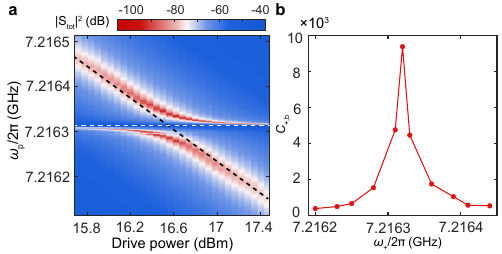}
\caption{\textbf{Anti-crossing in the normal mode spectrum. a} Output spectrum of the coupled upper-branch polariton and the mechanical mode versus the drive power.  
{Due to the magnon self-Kerr effect, the frequency of the polariton shifts by increasing the drive power (black dashed line).  The scattered mechanical sideband is fixed at $7.216313$ GHz (white dashed line).} {During the process of varying the drive power, the cross-Kerr effect causes a 1-kHz mechanical frequency shift, which has a negligible impact.} \textbf{b} Polaromechanical cooperativity $C_{\rm +,b}$ versus the frequency of the upper-branch polariton {$\omega_{+}$}. The cooperativity gets significantly improved when the polariton resonates with the mechanical sideband at the CPA frequency. }
\label{fig4}
\end{figure}

Another typical characteristics of the strong coupling is the anti-crossing, i.e., the level repulsion, in the frequency spectrum.  In Fig.~\ref{fig4}a, we fix the drive frequency at $\omega_{\rm{d}}/2\pi \,{=}\,7.205365$ GHz, and thereby the scattered mechanical sideband is at $\omega_{\rm{d}} \,{+}\, \omega_{\rm{b}} \,{=}\, 2\pi \times 7.216313$ GHz. By varying the drive power and utilizing the magnon self-Kerr effect, the frequency of the upper-branch polariton can be continuously adjusted.  When the polariton frequency is approaching and then passing across the mechanical sideband, the normal-mode spectrum shows an anti-crossing feature due to the polaromechanical  strong coupling.   It is worth noting that the mechanical sideband is set at the CPA frequency $\omega_{\rm CPA}$, at which the polariton has the smallest decay rate. {In addition, we observe a weaker anti-crossing signal beside the main one in Fig.~\ref{fig4}a, which results from the strong coupling with the adjacent mechanical mode.} Figure~\ref{fig4}b shows a significant enhancement of the polaromechanical cooperativity as the frequency of the upper-branch polariton approaches the CPA frequency.

\vspace{0.3cm}
\noindent\textbf{Conclusion and discussion}\\
\noindent
We have demonstrated {strong coupling} in a polaromechanical hybrid system, where the mechanical vibration mode is strongly coupled to the driven magnon polariton, whose decay rate is significantly reduced via realizing the coherent perfect absorption. We have measured the corresponding normal-mode splitting at the mechanical sideband, and a high polaromechanical cooperativity $C_{\rm +,b} = 9.40\times10^3$, which is {\it three orders of magnitude} higher than the previous experiments~\cite{Tang,Davis,Li2022}. This will significantly boost the mechanical cooling efficiency and lower the threshold for achieving phonon lasers in cavity magnomechanics~\cite{Davis}.

A quantum cooperativity $C_{\rm +,b}^{Q}\equiv C_{\rm +,b}/\bar{n}_b \gg 1$ ($\bar{n}_b$, the mean thermal phonon number~\cite{Painter20}) can be reached if placing the system at cryogenic temperatures {(see SI for estimation)}. This would enable various quantum applications, e.g., the protocols for preparing macroscopic quantum states of magnons and phonons~\cite{Jie18,Jie19,Jie19njp}, which are useful in testing the limits of quantum mechanics, and quantum entangled or squeezed states of microwave fields~\cite{Jie20,NSR}. 
The strong coupling that we report here can act as a foundation for building multifunctional HQSs involving more different quanta, e.g., by coupling the microwave cavity to superconducting qubits~\cite{Naka19} and the mechanical vibration to optical photons~\cite{OMrmp}. Our results can also be applied to YIG nano-structures~\cite{Heyroth19} or planar configurations~\cite{Xu21,YLi19,Hatanaka22}, which would enable the integration of magnonic devices on a chip and thus improve the functionality of the HQS to be used in quantum information processing or quantum sensing.

\vspace{0.5cm}
\noindent
\textbf{Methods}\\
The magnonic system shows an excellent tunability in its resonance frequency. In this work, we utilize two methods to adjust the frequency of the magnon mode. One is to adjust the external bias magnetic field via $\omega_{\rm m}=\gamma B_{0}$; the other is to use the magnon self-Kerr effect, reflected in the frequency of the driven magnon mode $\tilde{\omega}_{\rm{m}} \approx \omega_{\rm{m}}+2 K_{\rm{m}}|M|^{2}$~(see SI for details). Both methods can achieve the same effect. When the strong drive field is absent, we alter the external bias magnetic field to adjust the magnon frequency, as done in Fig.~\ref{fig2}a. 
When the drive field is applied, we alter the power of the drive field to change the magnon excitation number, which adjusts the magnon frequency due to the self-Kerr effect, as done in Fig.~\ref{fig4}a.

In the experiment, we measure a negative self-Kerr coefficient $ K_{\rm{m}}/2\pi=-7.4~\rm{nHz}$. The magnon mode experiences a negative frequency shift by increasing the drive power. To compensate for this effect, Fig.~\ref{fig3}b is measured under different initial frequencies of the upper-branch polariton. From the top to the bottom graph, we increase the external magnetic field $B_{0}$ to have a higher initial magnon frequency (without applying the drive), and thus a higher initial frequency of the upper-branch polariton. This compensates the negative frequency shift of the magnon mode when the drive is applied. {On the other hand, the cross-Kerr effect results in a frequency shift of the mechanical mode, and the magnomechanical coupling also leads to the modification of the CPA conditions, i.e., a shift of the CPA frequency (see Secs.~IV and VI of SI). In the measurements of Fig.~\ref{fig3}b, we adjust the drive frequency considering the combined effect of the above two factors to ensure that, in the process of measuring the output spectrum at various drive powers, the anti-Stokes mechanical sideband always resonates with the polariton at the CPA frequency.} 

\vspace{0.1cm}
\noindent
\textbf{Data availability}\\
\noindent The datasets used to generate the plots in the paper are available on Zenodo (https://zenodo.org/records/12732420). All other data that support the plots within this paper and other findings of this study are available from the corresponding authors upon reasonable request. \\



\vspace{0.1cm}
\noindent
\textbf{Acknowledgments}\\
The authors thank Liu Qiu and Yanhao Tang for useful discussions. This work was supported by National Key Research and Development Program of China (Grant Nos. 2022YFA1405200, 2024YFA1408900), National Natural Science Foundation of China (Grant Nos. 92265202, 11934010, 12174329, 12474365), the Innovation Program for Quantum Science and Technology (Grant No. 2021ZD0300200), Zhejiang Provincial Natural Science Foundation of China (Grant No. LR25A050001), and the Fundamental Research Funds for the Central Universities (No. 2021FZZX001-02).  \\

\vspace{0.1cm}
\noindent
\textbf{Author contributions} \\
R.C.S. performed the measurements and R.C.S. and X.Z. developed the theory. The measurement and theory related to the polariton-mechanics coupling is under the supervision of J.L. Y.M.S. and W.J.W. provided experimental support. R.C.S., J.L., Y.P.W. and J.Q.Y. analyzed the data. R.C.S., J.L. and J.Q.Y. wrote the manuscript with the input and comment from all co-authors. J.Q.Y. supervised the project. \\

\vspace{0.1cm}
\noindent
\textbf{Competing interests} \\
The authors declare no competing interests.\\

\vspace{0.1cm}
\noindent
\textbf{Additional information} \\
\noindent\textbf{Supplementary information} The online version contains supplementary material available at


\setcounter{figure}{0}
\setcounter{equation}{0}
\setcounter{table}{0}
\renewcommand\theequation{S\arabic{equation}}
\renewcommand\thefigure{S\arabic{figure}}
\renewcommand\thetable{S\arabic{table}}

\setcounter{section}{0}

\clearpage    
\onecolumngrid  

\section*{Supplemental  Information }

\maketitle

\subsection{System parameters}

	\begin{table}[htbp]
	\centering
	\begin{tabular}{| c | c | c | }
		\hline
		Quantity & Symbol & Value\\
		\hline
        \hline
		 Gyromagnetic ratio & $\gamma$ & $2\pi \times$2.8 MHz/Oe \\
		\hline
		 Frequency of the cavity $\rm{TE}_{\rm{102}}$ mode & $\omega_{\rm{a}}$ & $2\pi \times$7.213 GHz \\
		\hline
                 Frequency of the magnon mode & $\omega_{\rm{m}}$ & $\gamma B_0$ \\
		\hline
		 Frequency of the mechanical mode 1 & $\omega_{\rm{b,1}}$ & $2\pi \times$10.9565 MHz \\
        \hline
		 Frequency of the mechanical mode 2 & $\omega_{\rm{b,2}}$ & $2\pi \times$10.9511 MHz \\
		\hline
         Frequency of the mechanical mode 3 & $\omega_{\rm{b,3}}$ & $2\pi \times$10.9388 MHz \\
		\hline
                Intrinsic decay rate of the cavity mode & $\kappa_{\rm{int}}$ & $2\pi \times$1.98 MHz \\
              \hline
               External decay rate of the cavity mode & $\kappa_{\rm{1}}$ & $2\pi \times$1.22 MHz \\
                \hline
               External decay rate of the cavity mode & $\kappa_{\rm{2}}$ & $2\pi \times$1.24 MHz \\
	      \hline
               Damping rate of the magnon mode & $\kappa_{\rm{m}}$ & $2\pi \times$0.49 MHz \\
	       \hline
	        Damping rate of the mechanical mode 1 & $\kappa_{\rm{b,1}}$ & $2\pi \times$155 Hz \\
               \hline
	        Damping rate of the mechanical mode 2 & $\kappa_{\rm{b,2}}$ & $2\pi \times$230 Hz \\
          \hline
	        Damping rate of the mechanical mode 3 & $\kappa_{\rm{b,3}}$ & $2\pi \times$190 Hz \\
                \hline
                Cavity-magnon coupling strength & $g_{\rm{ma}}$ & $2\pi \times$6.63 MHz \\
		\hline
		 Bare magnomechanical coupling strength (mechanical mode 1) & $g_{\rm{mb,1}}$ & $2\pi \times$1.40 mHz \\
                \hline
		 Bare magnomechanical coupling strength (mechanical mode 2) & $g_{\rm{mb,2}}$ & $2\pi \times$0.29 mHz \\
          \hline
		 Bare magnomechanical coupling strength (mechanical mode 3) & $g_{\rm{mb,3}}$ & $2\pi \times$0.20 mHz \\
               \hline
                Magnon self-Kerr coefficient  & $K_{\rm{m}}$ & $-2\pi \times$7.4 nHz \\
		\hline
	\end{tabular}
	\caption{List of system parameters.}
	\label{tab:1}
\end{table}

\begin{figure}[h]
	\hskip-0.35cm\includegraphics[width=0.8\linewidth]{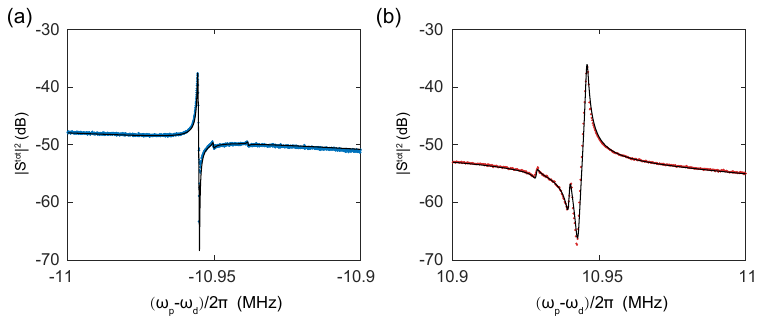}
	\caption{ (a) The output spectrum of the Stokes mechanical sideband at $\omega_{\rm S}=\omega_{\rm d}-\omega_{\rm b}$, with the drive power of 0.1 dBm. (b) The output spectrum of the anti-Stokes mechanical sideband at $\omega_{\rm AS}=\omega_{\rm d}+\omega_{\rm b}$, with the drive power of 16 dBm. The natural frequencies of the three mechanical modes are 10.9565, 10.9511, and 10.9388 MHz, respectively. }
	\label{figS1}
\end{figure}

Here we provide more details on the system parameters and setup used in the experiment, cf., Table \ref{tab:1}.  In the experiment, the whole cavity is placed in a static magnetic field $B_{0}$ generated by a high-precision tunable electromagnet. The diameter of the pole cap of the electromagnet is about 10 cm, which can provide a uniform magnetic field to magnetize the YIG sphere. The frequency of the magnon mode is determined by the external magnetic field via $\omega_{\rm m}=\gamma B_{0}$, with the gyromagnetic ratio $\gamma/2\pi=28~{\rm GHz/T}$. In the experiment, we apply a magnetic field about $B_{0}=258~{\rm mT}$, and the magnon mode is near resonance with the cavity mode. The drive field is provided by the KEYSIGHT Analog Signal Generator E8257D, which can generate a continuous microwave field with the maximum power of 25 dBm. The microwave input field is generated by the KEYSIGHT Vector Network Analyzer (VNA) N5232B, with the power of -10 dBm. The input field is attenuated by 32 dB via the power splitter, variable attenuator, variable phase shifter, and directional coupler, of which 20 dB is caused by the directional coupler.  The output spectrum is measured by the VNA, which has a dynamic range of 130 dB.  We observe three mechanical modes with close resonance frequencies $\omega_{\rm{b,j}}$ ($j=1,2,3$), as shown in Fig.~\ref{figS1}, among which we focus on the mechanical mode~1 that exhibits the strongest bare magnomechanical coupling strength $g_{\rm{mb,1}}$ and lowest damping rate $\kappa_{\rm{b,1}}$.  {This mechanical mode corresponds to a maximal mode overlap with the magnon mode and its dissipation is minimally influenced by the contact with the glass capillary~\cite{Tang}.}

\subsection{System Hamiltonian and magnon self-Kerr nonlinearity}

The system, i.e., a cavity magnomechanical (CMM) system, consists of a three-dimensional (3D) oxygen-free copper microwave cavity and a yttrium-iron-garnet (YIG) sphere. The YIG sphere is placed at the antinode of the magnetic field of the cavity $\rm{TE_{102}}$ mode, and the magnon mode couples to the cavity mode via the magnetic-dipole interaction. By applying an external bias magnetic field, a uniform magnon mode is excited in the YIG sphere at the frequency $\omega_{m}=\gamma B_{0}$, where $\gamma$ is the gyromagnetic ratio and $B_{0}$ is the strength of the external magnetic field.  
The magnon mode is driven by a microwave field loaded via a loop antenna.
The magnon excitations couple to the mechanical displacement (elastic strain) of the YIG sphere via the magnetostrictive interaction, which is a radiation-pressure-like dispersive coupling. When the drive field is strong, the magnon mode also exhibits a self-Kerr effect and a cross-Kerr effect with a mechanical mode~\cite{Li2022}. By including these effects, the Hamiltonian of the system reads
\begin{equation}\label{sup1}
\begin{aligned}
H/\hbar  = &\omega_{\rm{a}} a^{\dagger} a+\omega_{\rm{m}} m^{\dagger} m+\omega_{\rm{b}} b^{\dagger} b+g_{\rm {ma}} (a^{\dagger} m+a m^{\dagger})+g_{\rm {mb}} m^\dagger m  \left (b+ b^\dagger\right)+ K_{\rm{m}} m^{\dagger} m m^{\dagger} m  + K_{\rm{cross}} m^{\dagger} m b^{\dagger} b  \\
& +i\Omega_{\rm{d}}\left(m^{\dagger} e^{-i\omega_{\rm{d}} t}-m e^{i \omega_{\rm{d}} t}\right)+i\sqrt{2\kappa_{\rm{1}}}a^{\rm{in}}_{1}\left(a^{\dagger} e^{-i\omega_{\rm{1}} t}-a e^{i \omega_{\rm{1}} t}\right)+i\sqrt{2\kappa_{\rm{2}}}a^{\rm{in}}_{2}\left(a^{\dagger} e^{-i\omega_{\rm{2}} t}-a e^{i \omega_{\rm{2}} t}\right),
\end{aligned}
\end{equation}
where $\omega_{\rm{a}}$, $\omega_{\rm{m}}$ and $\omega_{\rm{b}}$ are the resonance frequencies of the cavity, magnon and mechanical modes, respectively; $a^{\dag}(a)$, $m^{\dag}(m)$ and $b^{\dag}(b)$ are their associated creation (annihilation) operators. $K_{\rm{m}}$ is the magnon self-Kerr coefficient; $K_{\rm{cross}}$ is the magnon-phonon cross-Kerr coefficient. In the experiment, the cross-Kerr effect is much weaker than the magnon self-Kerr effect~\cite{Li2022}, which mainly leads to the frequency shift of the  mechanical mode with the modified frequency $\tilde{\omega}_{\rm{b}} = \omega_{\rm{b}}+ K_{\rm{cross}}|M|^{2}$, and can be neglected in the calculation of the magnon excitation number. $\Omega_{\rm{d}}$ is the magnon-drive coupling strength associated with the drive field at frequency $\omega_{\rm{d}}$; $a^{\rm{in}}_{1(2)}$ denotes the input (probe) field at frequency $\omega_{1(2)}$, and $\kappa_{\rm{1(2)}}$ is the associated external decay rate of the cavity port 1(2) for loading the input field. In the experiment, the strength of the input fields is much smaller than that of the drive field. Thus the input fields can be treated as a perturbation to the system. The Heisenberg-Langevin equations of the system are then given by
\begin{equation}\label{sup2}
\begin{aligned}
&\frac{da}{dt}=- \left(i\omega_{\rm{a}}+ \kappa_{\rm{a}} \right) a-ig_{\rm{ma}} m, \\
&\frac{dm}{dt}=-\left[i \left(\omega_{\rm{m}}+2 K_{\rm{m}}m^{\dagger}m +K_{\rm{m}} \right) +ig_{\rm{mb}}(b+b^{\dagger})+ \kappa_{\rm{m}}\right] m -ig_{\rm{ma}} a+\Omega_{\rm{d}}e^{-i\omega_{\rm{d}} t}, \\
& \frac{db}{dt}=-\left(i\tilde{\omega}_{\rm{b}} + \kappa_{\rm{b}} \right)b -ig_{\rm{mb}}m^{\dagger}m,
\end{aligned}
\end{equation}
where $\kappa_{\rm{a}}$, $\kappa_{\rm{m}}$ and $\kappa_{\rm{b}}$ are the dissipation rates of the three modes. The cavity decay rate $\kappa_{\rm{a}}=\kappa_{\rm{int}}+\kappa_{1}+\kappa_{2}$ consists of three parts with $\kappa_{\rm{int}}$ being the intrinsic decay rate, and $\kappa_{\rm{1,2}}$ being the external decay rates due to the connection to the two cavity ports. By writing each mode operator $o$ as the sum of its steady-state average $O$ and a small quantum fluctuation $\delta o$, i.e., $a\equiv A+\delta a$, $m\equiv M+\delta m$, and $b\equiv B+\delta b$, we obtain the following equations for the classical averages {\it in the steady state} in the frame rotating at the drive frequency $\omega_{\rm{d}}$:
\begin{equation}\label{sup3}
\begin{aligned}
&\left( \Delta_{\rm{ad}}-i\kappa_{\rm{a}} \right) A+g_{\rm{ma}} M=0, \\
&\left[\Delta_{\rm{md}}+2 K_{\rm{m}}|M|^{2} +K_{\rm{m}}+g_{\rm{mb}}(B+B^{*})-i \kappa_{\rm{m}} \right] M +g_{\rm{ma}} A+i\Omega_{\rm{d}}=0, \\
&\left(\tilde{\omega}_{\rm{b}}-i\kappa_{\rm{b}} \right)B +g_{\rm{mb}}|M|^{2}=0,
\end{aligned}
\end{equation}
where $\Delta_{\rm{ad}}=\omega_{\rm{a}}-\omega_{\rm{d}}$, and $\Delta_{\rm{md}}=\omega_{\rm{m}}-\omega_{\rm{d}}$. Solving the first and third equations in Eq.~\eqref{sup3}, we obtain
\begin{equation}\label{sup4}
\begin{aligned}
& A=-\frac{g_{\rm{ma}}}{\Delta_{\rm{ad}}-i\kappa_{\rm{a}}} M, \\
&B =-\frac{g_{\rm{mb}}}{\tilde{\omega}_{\rm{b}}-i\kappa_{\rm{b}} }|M|^{2}.
\end{aligned}
\end{equation}
Substituting Eq.~\eqref{sup4} into the second equation of Eq.~\eqref{sup3}, we achieve
\begin{equation}\label{sup5}
\begin{aligned}
\left[\Delta_{\rm{md}}  \,{+}\, 2 K_{\rm{m}}|M|^{2} {+}\,K_{\rm{m}} {-} \,\eta_{\rm{b}} |M|^{2}- \eta_{\rm{a}}\Delta_{\rm{ad}} - i \left(\kappa_{\rm{m}} +\eta_{\rm{a}} \kappa_{\rm{a}} \right)\right] M =-i\Omega_{\rm{d}},
\end{aligned}
\end{equation}
where $\eta_{\rm{a}}= \frac{g_{\rm{ma}}^2} {\Delta_{\rm{ad}}^2+\kappa_{\rm{a}}^2}$ and $\eta_{\rm{b}}= \frac{2\tilde{\omega}_{\rm{b}}g_{\rm{mb}}^2} {\tilde{\omega}_{\rm{b}}^2+\kappa_{\rm{b}}^2}$. Multiplying Eq.~\eqref{sup5} with its complex conjugate, we obtain the following equation, which is a function of the magnon excitation number $|M|^{2}$:
\begin{equation}\label{sup6}
\begin{aligned}
\left[\left(\Delta_{\rm{md}}  \,{+}\, 2 K_{\rm{m}}|M|^{2} {+}\,K_{\rm{m}} {-} \,\eta_{\rm{b}} |M|^{2}- \eta_{\rm{a}}\Delta_{\rm{ad}} \right)^2+ \left(\kappa_{\rm{m}} +\eta_{\rm{a}} \kappa_{\rm{a}} \right)^2\right] |M|^{2} =cP_{\rm{d}},
\end{aligned}
\end{equation}
where $cP_{\rm{d}} \equiv \Omega_{\rm{d}}^2$, with $P_{\rm{d}}$ being the drive power and $c=\Omega_{\rm{d}}^2/P_{\rm{d}}$ being the fitting parameter characterizing the coupling strength between the drive power and the magnon mode, which is found to be $c/(2\pi)^2=1.03\times10^{30}~\rm{Hz^2/W}$ in the experiment.

For the YIG sphere (with the diameter of 0.25 mm) used in the experiment, the magnon self-Kerr coefficient is $ K_{\rm{m}}/2\pi=-7.4~\rm{nHz}$. Under the drive powers used in the experiment, the terms $K_{\rm{m}}$ and $\eta_{\rm{b}} |M|^{2}$ are much smaller than $2 K_{\rm{m}}|M|^{2}$. Hence, Eq.~\eqref{sup6} can be approximated as
\begin{equation}\label{sup7}
\begin{aligned}
\left[\left(\Delta_{\rm{md}}  \,{+}\, 2 K_{\rm{m}}|M|^{2} - \eta_{\rm{a}}\Delta_{\rm{ad}} \right)^2+ \left(\kappa_{\rm{m}} +\eta_{\rm{a}} \kappa_{\rm{a}} \right)^2\right] |M|^{2} =cP_{\rm{d}}.
\end{aligned}
\end{equation}
It is a cubic equation of the magnon excitation number $|M|^{2}$. There are three solutions of $|M|^{2}$, but only one stable solution exists for the drive powers and magnon-drive detunings used in the experiment.  The power is constrained to avoid the appearance of bistability~\cite{Li2022}. The large number of the magnon excitations via the driving field gives rise to a frequency shift of the magnon mode (the self-Kerr effect), which is reflected in the frequency of the {\it driven} magnon mode, $\tilde{\omega}_{\rm{m}} \simeq \omega_{\rm{m}}+2 K_{\rm{m}}|M|^{2}$, with $\omega_{\rm{m}}$ being the magnon frequency without applying the drive.

\subsection{Output spectra of the input fields}

To get the output spectra of the probe fields, we linearize the system Hamiltonian~\eqref{sup1}, which is valid under the drive powers used in the experiment. The output spectra can be conveniently obtained by including the strong pump effects into the linearized Langevin equations~\cite{Li2022}. We obtain the following linearized Langevin equations for the classical averages $A$, $M$, and $B$ in the frequency domain:
\begin{equation}\label{sup8}
\begin{aligned}
&-i\omega A=-\left ( i\Delta_{\rm{ad}}+\kappa_{\rm{a}} \right ) A -ig_{{\rm{ma}}} M +\sqrt{2\kappa_{\rm{1}}}a^{\rm{in}}_{1}\delta(\omega_{\rm{1}}-\omega_{\rm{d}}-\omega) +\sqrt{2\kappa_{\rm{2}}}a^{\rm{in}}_{2}\delta(\omega_{\rm{2}}-\omega_{\rm{d}}-\omega),  \\
&-i\omega M=-\left ( i\tilde{\Delta}_{\rm{md}}+\kappa_{\rm{m}} \right ) M -ig_{\rm{ma}} A -i\sqrt{2} G_{{\rm{mb}}} X,\\
&-i\omega X= \tilde{\omega}_{\rm{b}} P,\\
&-i\omega P= -\tilde{\omega}_{\rm{b}} X- 2\kappa_{\rm{b}} P - \sqrt{2} G_{{\rm{mb}}}^{*} M,
\end{aligned}
\end{equation}
where, for the mechanical mode, we define the steady-state average of the deformation displacement $X=(B+B^*)/\sqrt{2}$ and momentum $P=i (B^* - B)/\sqrt{2}$, and the effective magnomechanical coupling strength $G_{{\rm{mb}}} =g_{{\rm{mb}}}M$. The magnon-drive detuning $\tilde{\Delta}_{\rm{md}}=\tilde{\omega}_{\rm{m}}-\omega_{\rm{d}}$ includes the frequency shift due to the magnon self-Kerr effect.   Solving Eq.~\eqref{sup8} for the cavity field, we obtain
\begin{equation}\label{sup9}
\begin{aligned}
A(\omega) = \frac{ 1-2i|G_{\rm{mb}}|^{2}\chi_{\rm{b}}(\omega)\chi_{\rm{m}}(\omega) } {g_{\rm{ma}}^{2} \chi_{\rm{m}}(\omega) + \chi^{-1}_{\rm{a}}(\omega) \big( 1-2i|G_{\rm{mb}}|^{2}\chi_{\rm{b}}(\omega)\chi_{\rm{m}}(\omega) \big) } \left(\sqrt{2\kappa_{\rm{1}}}a^{\rm{in}}_{1}+\sqrt{2\kappa_{\rm{2}}}a^{\rm{in}}_{2} \right),
\end{aligned}
\end{equation}
where $\chi_{\rm{j}}(\omega)$ ($j=a,m,b$) are the natural susceptibilities of the cavity, magnon, and mechanical modes, respectively, i.e.,
\begin{equation}\label{sup10}
\begin{aligned}
\chi_{\rm{a}}(\omega)&=\frac{1}{i\left(\Delta_{\rm{ad}}-\omega \right)+\kappa_{\rm{a}}} ,\\
\chi_{\rm{m}}(\omega)&=\frac{1}{i\left(\tilde{\Delta}_{\rm{md}}-\omega \right)+\kappa_{\rm{m}}}, \\
\chi_{\rm{b}}(\omega)&=\frac{\tilde{\omega}_{\rm{b}}}{\tilde{\omega}_{\rm{b}}^{2}-\omega^{2}-2i\kappa_{\rm{b}}\omega }.
\end{aligned}
\end{equation}

The output fields of the microwave cavity can then be obtained by using the input-output relation $A^{\rm{out}}_{i}=\sqrt{2\kappa_{\rm{i}}}A-a^{\rm{in}}_{i}$ ($i=1,2$)~\cite{Collett1985}, which are
\begin{equation}\label{sup11}
\begin{aligned}
A^{\rm{out}}_{1}&= r_{1}a^{\rm{in}}_{1}+t_{2}a^{\rm{in}}_{2},\\
A^{\rm{out}}_{2}&= t_{1}a^{\rm{in}}_{1} + r_{2}a^{\rm{in}}_{2}.
\end{aligned}
\end{equation}
Here, $r_{i}$ ($t_{i}$) is the reflection (transmission) of the port $i$. Such a two-channel system leads to a two-dimension scattering matrix $S(\omega)$, which relates two input fields to the corresponding two output fields via $(A^{\rm{out}}_{1}, A^{\rm{out}}_{2})^{\rm T}=S(\omega) (a^{\rm{in}}_{1}, a^{\rm{in}}_{2})^{\rm T}$, where $S(\omega)= \left(
                                \begin{array}{cc}
                                  r_{1} & t_{2} \\
                                  t_{1} & r_{2} \\
                                \end{array}
                              \right)
$, with
\begin{equation}\label{sup12}
\begin{aligned}
r_{i}= \frac{2\kappa_{\rm{i}} \left(1-2i|G_{\rm{mb}}|^{2}\chi_{\rm{b}}(\omega)\chi_{\rm{m}}(\omega) \right)} {g_{\rm{ma}}^{2} \chi_{\rm{m}}(\omega) + \chi^{-1}_{\rm{a}}(\omega) \big( 1-2i|G_{\rm{mb}}|^{2}\chi_{\rm{b}}(\omega)\chi_{\rm{m}}(\omega) \big) } -1,\\
t_{1}=t_{2}= \frac{2\sqrt{\kappa_{\rm{1}} \kappa_{\rm{2}} } \left(1-2i|G_{\rm{mb}}|^{2}\chi_{\rm{b}}(\omega)\chi_{\rm{m}}(\omega) \right)} {g_{\rm{ma}}^{2} \chi_{\rm{m}}(\omega) + \chi^{-1}_{\rm{a}}(\omega) \big( 1-2i|G_{\rm{mb}}|^{2}\chi_{\rm{b}}(\omega)\chi_{\rm{m}}(\omega) \big) } .
\end{aligned}
\end{equation}

Since the two input fields are generated by splitting the microwave signal into two beams using the power splitter, the two input fields are of the same frequency, $\omega_{1}=\omega_{2}$. Therefore, the two input fields can be written in the form of $a^{\rm{in}}_{2}=\sqrt{q}e^{-i\Delta\phi}a^{\rm{in}}_{1}$, with the phase difference $\Delta\phi$ and the power ratio $q$. The output spectra are then given by
\begin{equation}\label{sup13}
\begin{aligned}
|S^{\rm{tot}}_{1}|^{2}(\omega)& \equiv \left|\frac{A^{\rm{out}}_{1}}{a^{\rm{in}}_{1}}\right|^2\\
&= \left|r_{1}+ \sqrt{q}e^{-i\Delta\phi}t_{2} \right|^2,\\
|S^{\rm{tot}}_{2}|^{2}(\omega)& \equiv \left|\frac{A^{\rm{out}}_{2}}{a^{\rm{in}}_{2}}\right|^2\\
&= \left|r_{2}+ e^{i\Delta\phi}t_{1}/\sqrt{q} \right|^2.
\end{aligned}
\end{equation}
For the particular case of $q=\kappa_{\rm{2}}/\kappa_{\rm{1}}$, and $\Delta\phi=2n\pi$ ($n=0, 1, 2, \cdots$), we have $|S^{\rm{tot}}_{1}|^{2}(\omega)=|S^{\rm{tot}}_{2}|^{2}(\omega) \equiv |S^{\rm{tot}}|^{2}(\omega)$, which allows us to simultaneously tune the two output fields to be zero.

\subsection{Coherent perfect absorption for the cavity magnomechanical system}

Here, we derive the coherent perfect absorption (CPA) conditions for the CMM system.  By setting $|S^{\rm{tot}}|^{2}=0$~\cite{Chong2010}, which yields $a^{\rm{in}}_{1(2)} =\sqrt{2\kappa_{1(2)}}A$, and solving the above equation, we obtain the following conditions for realizing the CPA of the CMM system, which are $q=\kappa_{\rm{2}}/\kappa_{\rm{1}}$, $\Delta\phi=2n\pi$, and
\begin{equation}\label{sup14}
\begin{aligned}
\left(\kappa_{\rm{2}}+\kappa_{\rm{1}}- \kappa_{\rm{int}}\right)\rm{Im}[\Xi]+\eta_{\rm{m}}\left(\tilde{\Delta}_{\rm{md}}-\omega\right)&= \left(\Delta_{\rm{ad}}-\omega\right)\left(1+\rm{Re}[\Xi]\right),\\
\eta_{\rm{m}} \kappa_{\rm{m}} -  \left(\Delta_{\rm{ad}}-\omega\right) \rm{Im}[\Xi]&=\left(\kappa_{\rm{2}}+\kappa_{\rm{1}}- \kappa_{\rm{int}}\right) \left(1+\rm{Re}[\Xi]\right),
\end{aligned}
\end{equation}
where $\Xi=-2i|G_{\rm{mb}}|^{2}\chi_{\rm{b}}(\omega)\chi_{\rm{m}}(\omega)\equiv\rm{Re}[\Xi]+ i \rm{Im}[\Xi]$, and $\eta_{\rm{m}}=\frac{g_{\rm{ma}}^{2}}{\left(\tilde{\Delta}_{\rm{md}}-\omega\right)^{2}+\kappa_{\rm{m}}^{2}}$. The terms containing $\Xi$ can be considered as the modification on the CPA conditions due to the coupling with the mechanical mode, as seen from the fact that $\Xi=0$ when $G_{\rm{mb}}=0$.  For relatively small drive powers used in the experiment, $2|G_{\rm{mb}}|^{2} \ll \chi_{\rm{b}}^{-1}(\omega)\chi_{\rm{m}}^{-1}(\omega)$, such that $|\Xi|\ll1$. Consequently, the terms of $\Xi$ in Eq.~(\ref{sup14}) can be neglected (the mechanical modification on the CPA is negligible), and the CPA conditions are reduced to those for the cavity-magnon bipartite system, which are
\begin{equation}\label{sup15}
\begin{aligned}
\eta_{\rm{m}} \left(\tilde{\Delta}_{\rm{md}}-\omega\right)&= \Delta_{\rm{ad}}-\omega,\\
\kappa_{\rm{2}}+\kappa_{\rm{1}}- \kappa_{\rm{int}}&=\eta_{\rm{m}} \kappa_{\rm{m}}.
\end{aligned}
\end{equation}
With sufficiently strong powers to have the polariton-mechanics normal mode splitting (cf. Fig.~3 in the main text), the coupling to the mechanical mode will significantly modify the CPA conditions for the cavity-magnon system only, i.e., Eq.~\eqref{sup14}. In this case, one has to resort to the exact CPA conditions Eq.~\eqref{sup14} for the tripartite CMM system.

\begin{figure}[h]
	\hskip-0.35cm\includegraphics[width=0.8\linewidth]{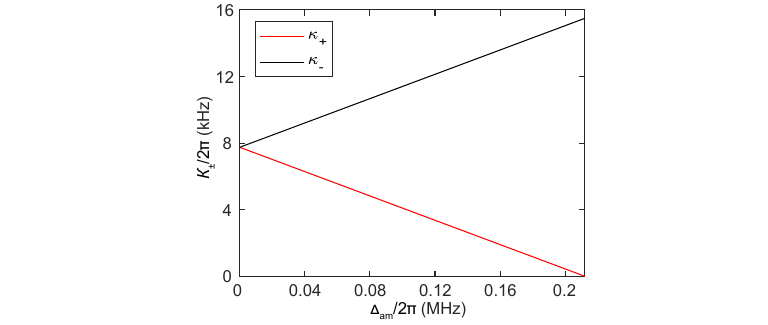}
	\caption{The decay rates of the CMPs $\kappa_{\pm}$ versus the cavity-magnon detuning $\Delta_{\rm am}$, for $\kappa_{\rm{int}}/2\pi=1.98~\rm{MHz}$, $\kappa_{\rm{1}}/2\pi=1.22~\rm{MHz}$, $\kappa_{\rm{2}}/2\pi=1.24~\rm{MHz}$, and $\kappa_{\rm{m}}/2\pi=0.49~\rm{MHz}$. }
	\label{figS2}
\end{figure}

In what follows, we derive the decay rates of the cavity-magnon polaritons (CMPs) under the CPA. This allows us to see clearly how the CPA modifies the CMP decay rates.
For simplicity, we limit our discussion to the situation where the mechanical modification on the CPA conditions is negligible. This can provide us analytical yet simple expressions of the decay rates $\kappa_{\pm}$ under the CPA.
Under the CPA, we have $a^{\rm{in}}_{1}=\sqrt{ 2\kappa_{\rm{1}}} A$ and $a^{\rm{in}}_{2}=\sqrt{ 2\kappa_{\rm{2}}} A$. By substituting them into Eq.~\eqref{sup8}, we obtain the following equations for the cavity-magnon system only:
\begin{equation}\label{sup16}
\begin{aligned}
-i\omega A&=-\left ( i\Delta_{\rm{ad}}+\kappa_{\rm{a}} \right ) A -ig_{{\rm{ma}}} M +2\kappa_{\rm{1}}A+2\kappa_{\rm{2}}A, \\
&\equiv - i\Delta_{\rm{ad}}A +\kappa_{\rm{a}}^{'} A -ig_{{\rm{ma}}} M,\\
-i\omega M&=-\left ( i\tilde{\Delta}_{\rm{md}}+\kappa_{\rm{m}} \right ) M -ig_{\rm{ma}} A.
\end{aligned}
\end{equation}
As clearly seen, the cavity mode is compensated to be a gain mode, with an {\it effective} gain rate $\kappa_{\rm{a}}^{'}=\kappa_{\rm{1}}+\kappa_{\rm{2}}-\kappa_{\rm{int}}$. From Eq.~\eqref{sup16}, we can extract the effective non-Hermitian Hamiltonian of the system
\begin{equation}\label{sup17}
\begin{aligned}
H_{\rm{ma}}/\hbar = (\omega_{\rm{a}}+i\kappa_{\rm{a}}^{'}) a^{\dagger} a+\left(\tilde{\omega}_{\rm{m}}-i \kappa_{\rm{m}}\right) m^{\dagger} m+g_{\rm {ma}} (a^{\dagger} m+a m^{\dagger}),
\end{aligned}
\end{equation}
which yields the following frequencies $\omega_{\pm}$ and decay rates $\kappa_{\pm}$ of the two CMPs:
\begin{equation}\label{sup18}
\begin{aligned}
\omega_{\pm}-i\kappa_{\pm}=\frac{\omega_{\rm{a}}+\tilde{\omega}_{\rm{m}}}{2}-i\frac{-\kappa_{\rm{a}}^{'}+\kappa_{\rm{m}}}{2} \pm \sqrt{g_{\rm{ma}}^2 +\left( \frac{\omega_{\rm{a}}-\tilde{\omega}_{\rm{m}}}{2}+i\frac{\kappa_{\rm{a}}^{'}+\kappa_{\rm{m}}}{2} \right)^2}.
\end{aligned}
\end{equation}
For the case where the cavity and the magnon mode are resonant, $\omega_{\rm{a}} = \tilde{\omega}_{\rm{m}}$, we have
\begin{equation}\label{sup19}
\begin{aligned}
\omega_{\pm}&=\omega_{\rm{a}} \pm \sqrt{g_{\rm{ma}}^2 -\left( \frac{\kappa_{\rm{a}}^{'}+\kappa_{\rm{m}}}{2} \right)^2},\\
\kappa_{\pm}&=\frac{-\kappa_{\rm{a}}^{'}+\kappa_{\rm{m}}}{2} .
\end{aligned}
\end{equation}
When the cavity gain rate and the magnon decay rate are further balanced, i.e., $ \kappa_{\rm{a}}^{'}= \kappa_{\rm{m}}$, the decay rates of the two CMPs reduce to zero, $\kappa_{\pm}=0$. In the output spectrum, the CPA occurs at the frequencies of the CMPs, i.e., $\omega=\omega_{\pm}-\omega_{\rm{d}}$ (in the frame rotating at $\omega_{\rm{d}}$), which leads to $\eta_{\rm{m}}=1$. The CPA conditions in Eq.~(\ref{sup15}) then become
\begin{equation}\label{sup20}
\begin{aligned}
\tilde{\Delta}_{\rm{md}}&= \Delta_{\rm{ad}},\\
 \kappa_{\rm{m}} &= \kappa_{\rm{a}}^{'}.
\end{aligned}
\end{equation}
Clearly, Eq.~(\ref{sup20}) agrees with Eq.~(\ref{sup19}). In the experiment, the external decay rate $\kappa_{1(2)}$ depends on the length of the pin inserted into the cavity, i.e., $\delta l_{1(2)}$. By tuning  $\delta l_{1}$ and $\delta l_{2}$, $\kappa_{1}+\kappa_{2}$ can be adjusted to be close to $\kappa_{\rm{int}}+\kappa_{\rm{m}}$. Limited by the regulation precision, we achieve the minimum symmetric decay rates $\kappa_{\pm}/2\pi=7.75~\rm{kHz}$, corresponding to $\kappa_{\rm{1}}/2\pi=1.22~\rm{MHz}$ and $\kappa_{\rm{2}}/2\pi=1.24~\rm{MHz}$. To further reduce the decay rate, we tune the magnon frequency to be off-resonant with the cavity. In this case, the decay rates of the two CMPs are no longer equal: One is reduced accompanied with the increase of the other, as shown in Fig.~\ref{figS2}. In this way, we further reduce the decay rate of the upper-branch polariton to $\kappa_{+}/2\pi= 0.78$ kHz, limited to the control accuracy.

\subsection{Coupling between polaritons and the mechanical mode}

The strong coupling between the cavity mode and the magnon mode leads to the hybridization of the two modes, forming two CMPs. Therefore, it is more convenient to describe the strongly-coupled cavity-magnon system with the polariton operators
\begin{equation}\label{sup21}
\begin{aligned}
p_{+}&=a\cos\theta + m\sin \theta ,\\
p_{-}&=-a\sin \theta + m\cos \theta ,
\end{aligned}
\end{equation}
where $p_{+(-)}$ denotes the annihilation operator of the upper (lower)-branch polariton mode, and $\theta= \frac{1}{2} \arctan \left( \frac{2g_{\rm {ma}}}{\omega_{\rm{a}}-\tilde{\omega}_{\rm{m}}} \right)$ determines the weight of the magnon (cavity) mode in the two polaritons. Here, we include the magnon self-Kerr induced frequency shift in the magnon frequency $\tilde{\omega}_{\rm m}$.  In this way, the Hamiltonian of the CMM system~\eqref{sup1} (neglecting the weak probe field) can then be described using the two polariton operators $p_{\pm}$ and the operator $b$ of the mechanical mode. The Hamiltonian is intrinsically nonlinear due to the dispersive magnetostrictive interaction, but it can be linearized under a not too weak drive field yielding $| \langle p_+ \rangle|,\, | \langle p_- \rangle| \gg 1$. The linearized Hamiltonian is given by
\begin{equation}\label{sup22}
\begin{aligned}
H/\hbar = \omega_{\rm{+}} p^{\dagger}_{+} p_{+}+\omega_{\rm{-}} p^{\dagger}_{-} p_{-}+{\omega}_{\rm{b}}b^{\dagger} b+\left(G_{+} p_{+}^\dagger+ G_{+}^{*} p_{+} \right) x +\left(G_{-} p_{-}^\dagger+ G_{-}^{*} p_{-} \right) x,
\end{aligned}
\end{equation}
where we have assumed $\tilde{\omega}_{\rm{b}} \simeq {\omega}_{\rm{b}}$ due to the small mechanical frequency shift in the experiment (Sec.~VI). $x= \left(b+ b^\dagger\right)/\sqrt{2}$ is the operator of the mechanical displacement, and $G_{+(-)}$ is the effective coupling strength between the upper (lower)-branch polariton and the mechanical mode, of which the expressions are
\begin{equation}\label{sup23}
\begin{aligned}
G_{+} &=\sqrt{2}g_{\rm{mb}}P_{+} \sin^2 \theta+\sqrt{2}g_{\rm{mb}}P_{-} \sin \theta \cos \theta=\sqrt{2}g_{\rm{mb}} M\sin \theta,\\
G_{-} &= \sqrt{2}g_{\rm{mb}}P_{-} \cos^2 \theta+\sqrt{2}g_{\rm{mb}}P_{+} \sin \theta \cos \theta=\sqrt{2}g_{\rm{mb}} M\cos \theta,
\end{aligned}
\end{equation}
with $P_{+(-)}=\left \langle p_{+(-)} \right \rangle$. The frequencies of the two polariton modes $ \omega_{\pm}$ are given in Eq.~(\ref{sup18}). The linearized polaromechanical Hamiltonian~\eqref{sup22} indicates that the interaction between each polariton and the mechanical mode is the ``quadrature-quadrature"-type coupling, which is analogous to the linearized optomechanical Hamiltonian~\cite{MAreview}. This implies that similar magnomechanical backactions as the optomechanical ones will be present, such as the cooling (amplification) of the mechanical mode when a red (blue)-detuned microwave field is used to drive the CMP to activate the anti-Stokes (Stokes) scattering~\cite{Li2022}.

\begin{figure}[h]
	\hskip-0.35cm\includegraphics[width=0.8\linewidth]{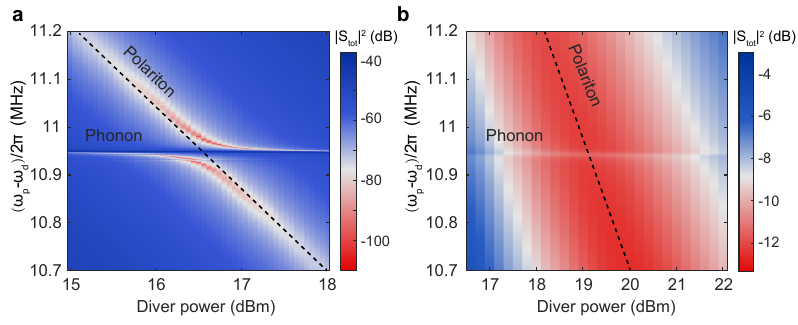}
	\caption{Output spectrum versus the drive power when the CPA conditions are satisfied in \textbf{a}; and not satisfied in \textbf{b}. An anti-crossing occurs in \textbf{a}, signifying the polaromechanical strong coupling, due to the significantly reduced polariton decay rate. }
	\label{figS3}
\end{figure}

{In what follows, to provide a definitive and intuitive proof of the role of the CPA in significantly reducing the polariton decay rate and thus observing the polaromechanical normal-mode splitting, we show in Fig.~\ref{figS3} the measurement of the output spectrum in two situations: the system is operated at the CPA conditions (Fig.~\ref{figS3}a); and away from the CPA conditions (Fig.~\ref{figS3}b). Clearly, the normal-mode splitting occurs only when the CPA conditions are satisfied. Note that in Fig.~\ref{figS3}a the CPA is achieved following the approach described in the main text, while the CPA conditions are not met in Fig.~\ref{figS3}b, which is obtained by sending a microwave input field and directly measuring its reflection at the input port.}

\subsection{Frequency shift of the mechanical mode}

\begin{figure}[h]
	\hskip-0.35cm\includegraphics[width=0.8\linewidth]{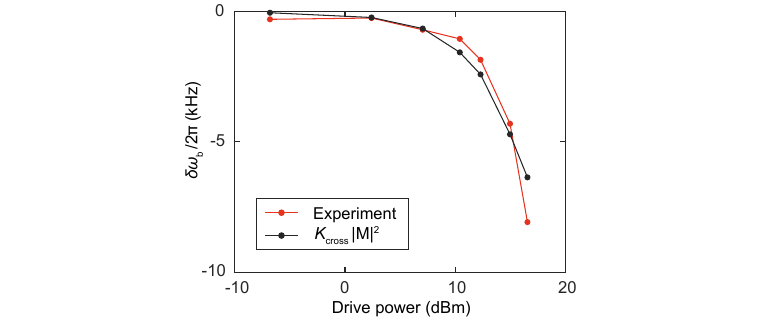}
	\caption{The frequency shifts of the mechanical mode under the drive powers used in Fig. 3(b) of the main text. Red dots are the experimental data and black dots are the frequency shifts induced by the cross-Kerr effect, evaluated via $\delta\omega_{\rm{b}}= K_{\rm{cross}} |M|^{2}$, with $K_{\rm{cross}}/2\pi \,\,{=}\,{-}6.8~\rm{pHz}$.}
	\label{figS4}
\end{figure}

In the experiment, we observe the frequency shift of the mechanical mode, which is mainly caused by the cross-Kerr effect between the magnon mode and the mechanical mode~\cite{Li2022}, with $\delta\omega_{\rm{b}}= K_{\rm{cross}} |M|^{2}$. The spring effect can also lead to the mechanical frequency shift, but its contribution is much smaller than the cross-Kerr effect. By fitting the experimental data, we obtain $K_{\rm{cross}}/2\pi \,\,{=}\,{-}6.8~\rm{pHz}$. In Fig.~\ref{figS4}, the red dots are the experimental data of the frequency shifts of the mechanical mode in Fig. 3(b) of the main text. In Fig. 3(b), each graph has a different initial magnon frequency (cf. {\bf Methods} in the main text).  
Therefore, the magnon excitation number $|M|^{2}$ of the black fitting dots in Fig.~\ref{figS4} are calculated under different initial magnon frequencies.  When the drive power becomes high, there is discrepancy between the theory and the experiment data, which could be due to the thermal effect caused by the strong power, which is not included in our model.


\subsection{Estimation of quantum cooperativity at cryogenic temperatures}


{The quantum cooperativity $C_{\rm +,b}^{Q}\equiv C_{\rm +,b}/\bar{n}_b$ greater than 1 is typically considered as a prerequisite to perform coherent quantum operations. For the mechanical mode of $\sim$10 MHz in our experiment, $\bar{n}_b \approx 20$ for a low bath temperature of $\sim$10 mK. This leads to a maximum quantum cooperativity $C_{\rm +,b}^{Q} = 470$ if using the maximum cooperativity  $C_{\rm +,b} = 9.4 \times 10^3$ achieved in our room-temperature experiment.  Thanks to this high cooperativity, the quantum cooperativity greater than 1 can still be achieved for the bath temperature being up to 4.5 K. 
}

\end{document}